\documentclass[twocolumn,showpacs,superscriptaddress,amssymb,10pt,pra,floatfix]{revtex4}

\usepackage{graphicx}% Include figure files
\usepackage{dcolumn}% Align table columns on decimal point
\usepackage{bm}% bold math
\usepackage{epsfig}
\usepackage{color}
\usepackage{longtable}

\def\twobytwo#1#2#3#4{             \begin{array}{cc}
                                   #1 & #2   \\[0.4cm]
                                   #3 & #4
                                   \end{array}    }

\begin{document}

\preprint{}
%
%
%-----------------------------------------Title---------------------------------------
%
%
\title{Scattering of twisted relativistic electrons by atoms}
%
%
%-----------------------------------------Author--------------------------------------
%
%

\author{V.~Serbo}
\affiliation{Novosibirsk State University, RUS--630090, Novosibirsk, Russia}
\affiliation{Sobolev Institute of Mathematics, RUS--630090, Novosibirsk, Russia}

\author{I.~P.~Ivanov}
\affiliation{CFTP, Instituto Superior Tecnico, University of Lisbon, Avenida Rovisco Pais, 1 1049--001 Lisbon, Portugal}

\author{S.~Fritzsche}
\affiliation{Helmholtz-Institut Jena, D--07743 Jena, Germany}
\affiliation{Theoretisch--Physikalisches Institut, Friedrich--Schiller--Universit\"at Jena, D--07743 Jena, Germany}

\author{D.~Seipt}
\affiliation{Helmholtz-Institut Jena, D--07743 Jena, Germany}

\author{A.~Surzhykov}
\affiliation{Helmholtz-Institut Jena, D--07743 Jena, Germany}

\date{\today \\[0.3cm]}

%
%
%-----------------------------------------Abstract--------------------------------------
%
%

\begin{abstract}
The Mott scattering of high--energetic twisted electrons by atoms is investigated within the framework of the first Born approximation and Dirac's relativistic equation. Special emphasis is placed on the angular distribution and longitudinal polarization of the scattered electrons. In order to evaluate these angular and polarization properties we consider two experimental setups in which the twisted electron beam collides with either a single well--localized atom or macroscopic atomic target. Detailed relativistic calculations have been performed for both setups and for the electrons with kinetic energy from 10~keV to 1000~keV. The results of these calculations indicate that the emission pattern and polarization of outgoing electrons differ significantly from the scattering of plane--wave electrons and can be very sensitive to the parameters of the incident twisted beam. In particular, it is shown that the angular-- and polarization--sensitive Mott measurements may reveal valuable information about, both the transverse and longitudinal components of the linear momentum and the projection of the total angular momentum of twisted electron states. Thus, the Mott scattering emerges as a diagnostic tool for the relativistic vortex beams.
\end{abstract}

\pacs{03.65.Nk, 03.65.Pm, 34.80.Bm}
\maketitle

%
%
%------------------------------------------------------  Introduction  ---------------------------------------------------------------
%
%
\section{Introduction}

Beams of photons and charged particles that carry a \textit{non--zero} projection of the orbital angular momentum (OAM) upon their propagation direction attract considerable attention in both, fundamental and applied research. During the last two decades, for example, a large number of  experiments have been performed with \textit{optical} twisted (or vortex) beams \cite{MoT07,FrA08,YaP11,ToT11}. Moreover, very recently the production and manipulation of the twisted \textit{electrons} have become feasible \cite{UcT10,VeT10,McM11,Ver11}. These vortex electrons, whose OAM projection may be as high as $\hbar m = 200 \hbar$ \cite{GrG15}, serve today as a valuable tool for probing the structural, electronic and magnetic properties with atomic resolution. The engineering of the sub--nanometer sized vortices opens also new possibilities for studying the shell structure and dynamics of individual atoms \cite{Ver11,ScV10}.

Most applications of twisted electron beams to the study of (material) structure rely on the knowledge about the electron \textit{scattering} by target atoms or ions. Accurate theoretical description of the basic scattering processes is required, therefore, which properly accounts for the details of the electron--atom interaction and the vortex structure of the incident beams. Moreover, such a theoretical analysis has to be necessarily performed within the \textit{relativistic} framework. This is due to the fact that the electron vortices are currently produced in transmission electron microscopes \cite{UcT10,McM11} with the typical kinetic energy of 300~keV, corresponding to the velocity of about 80~\% of the speed of light.
Even much higher energies up to 50~MeV are likely to be achieved in an experiment at the Thomas Jefferson National Accelerator Facility where the twisted electrons, created with the help of conventional holographic mask technique, are to be injected into the linear accelerator \cite{Dut14}. The scattering of these MeV--beams on solid targets, which will be observed to verify the vortex nature of the accelerated electrons, will proceed in an ultra--relativistic regime and definitely requires the proper theoretical treatment based on Dirac's equation.

Despite the definite need for the \textit{relativistic} analysis of the twisted electron scattering by atomic targets, this field of research has barely started to develop. Up to the present, most of the collision studies with the vortex beams have been performed in the non--relativistic framework \cite{BlB07,VaB13,MaH14}. In contrast, the twisted electron solutions of the Dirac equation have been applied either in free space \cite{BlD11} or in the presence of strong laser fields \cite{Kar12,HaM14,SeS14}, but not in the scattering kinematics. Here we close this gap and investigate the Mott scattering of \textit{relativistic} twisted electrons. In our study, we pay special attention to the angular distribution and polarization of the outgoing electrons. In order to calculate these (angular and polarization) properties we made use of the first Born approximation and the free--particle solutions of the Dirac equation. Moreover, we approximated the atomic potential by a sum of Yukawa terms; an approach that is justified for the description of the relativistic electron--atom collisions \cite{Sal91}. For such a potential, we evaluated in Section \ref{sec_theory} the first--Born scattering amplitudes for both, initial plane--wave and twisted Bessel electrons. With the help of these amplitudes we were able to derive then the angular distribution and polarization of the outgoing electrons. For the incident vortex beam, moreover, two experimental scenarios have been considered in which the target is (i) just a single well--localized atom, or (ii) a macroscopic ensemble of randomly distributed atoms. The relativistic calculations have been performed for both scenarios as well as for the neutral hydrogen and iron targets, and incident electron energies in the range 10~$\le T_e \le$~1000~keV. These calculations are discussed in Section~\ref{sec_computations} and clearly indicate that the angular distribution and polarization of the scattered electrons are sensitive not only to the (ratio of) transverse and longitudinal linear momenta but also to the projection of the total angular momentum (TAM) of the incident vortex beam. Even though these results have been obtained within the first Born approximation, which has a rather limited accuracy for the description of (realistic) electron--atom collisions, they allowed us to elucidate the main effects arising in the relativistic Mott scattering of twisted beams. Our study, therefore, can be viewed as a starting point for more elaborate calculations including spin--orbit and spin--spin interactions as well as higher terms of the perturbative expansion.

Relativistic units ($\hbar = 1$ and $c = 1$) are used throughout the paper.

%
%
%------------------------------------------------------  Theory  ---------------------------------------------------------------
%
%
\section{Theoretical background}
\label{sec_theory}

\subsection{Plane--wave electron scattering}
\label{subsec_plane_wave}

The (elastic) Mott scattering of plane--wave electrons has been worked out long ago and discussed in many textbooks. In the present paper, therefore, we will restrict ourselves to a rather short compilation of the basic relations, which will be used later to treat the twisted electron scattering.

\subsubsection{Evaluation of the transition amplitude}
\label{subsubsec_amplitude_plane}

The total Mott cross section as well as the angular distribution and polarization of outgoing (scattered) electrons are usually expressed in terms of the \textit{scattering amplitude}. Within the framework of the first Born approximation, this amplitude can be written as:
\begin{equation}
   \label{eq_scattering_amplitude_general}
   f_{\lambda \lambda'}({\bm p}, {\bm p}') = - \frac{1}{4 \pi} \int{\psi_{{\bm p}' \lambda'}^\dag({\bm r})
   \, V({\bm r}) \, \psi_{{\bm p} \lambda}({\bm r}) \, {\rm d}{\bm r}} \, ,
\end{equation}
where $V({\bm r})$ is the electron--atom interaction potential. In Eq.~(\ref{eq_scattering_amplitude_general}), moreover, the wave--functions $\psi_{{\bm p} \lambda}({\bm r})$ and $\psi_{{\bm p}' \lambda'}({\bm r})$ describe a \textit{free} electron in its initial and final states, characterized by the momenta ${\bm p}$ and ${\bm p}'$ and helicities $\lambda$ and $\lambda'$. These functions are plane--wave solutions of the Dirac equation and read:
\begin{equation}
   \label{eq_plane_wave_functions}
   \psi_{{\bm p} \lambda}({\bm r}) = u_{{\bm p} \lambda} \, {\rm e}^{i {\bm p} {\bm r}} \, , \: \: \: \:
   \psi_{{\bm p}' \lambda'}({\bm r}) = u_{{\bm p}' \lambda'} \, {\rm e}^{i {\bm p}' {\bm r}} \, ,
\end{equation}
where $u_{{\bm p} \lambda}$ (and $u_{{\bm p}' \lambda'}$) is the Dirac bi--spinor \cite{BeL82,EiM95}:
\begin{eqnarray}
   \label{eq_Dirac_bispinor}
   u_{{\bm p} \lambda} = \left( \begin{array}{c}
                                   \sqrt{\varepsilon + m_e} \, w^{(\lambda)}({\bm n}) \\[0.4cm]
                                   2 \lambda \sqrt{\varepsilon - m_e} \, w^{(\lambda)}({\bm n})
                                   \end{array}
                          \right) \, .
\end{eqnarray}
In this expression, $\varepsilon = \sqrt{m_e^2 + p^2}$ and ${\bm n} = {\bm p}/p$ is the total energy and the propagation direction of an electron, and spinor $w^{(\lambda)}({\bm n})$ is the eigenfunction of the helicity operator $\Lambda({\bm n}) = \hat{{\bm \sigma}} {\bm n} /2$,
\begin{equation}
   \label{eq_eigenstate_helicity_operator}
   \Lambda({\bm n}) w^{(\lambda)}({\bm n}) \equiv \frac{\hat{{\bm \sigma}} {\bm n}}{2} \, w^{(\lambda)}({\bm n}) = \lambda w^{(\lambda)}({\bm n}) \, ,
\end{equation}
with $\hat{\bm \sigma}$ being the vector of Pauli matrices. As seen from this expression, the helicity $\lambda = \pm 1/2$ is the projection of the electron spin onto its own direction of propagation.

For the further evaluation of the scattering amplitude (\ref{eq_scattering_amplitude_general}) one needs to find the explicit form of the spinor $w^{(\lambda)}({\bm n})$. To achieve this goal, let us consider first the electron propagation along the quantization ($z$--) axis of the overall system. In this case ${\bm n} = {\bm e}_z$ and the Eq.~(\ref{eq_eigenstate_helicity_operator}) simplifies to:
\begin{equation}
   \label{eq_eigenstate_helicity_operator_z_axis}
   \frac{\hat{\sigma}_z}{2} \, w^{(\sigma)}({\bm e}_z) \equiv \hat{s}_z w^{(\sigma)}({\bm e}_z) = \sigma w^{(\sigma)}({\bm e}_z) \, ,
\end{equation}
thus indicating that $w^{(\lambda)}({\bm e}_z)$ is just the standard Pauli spinor:
\begin{eqnarray}
   \label{eq_Dirac_spinor}
   w^{(1/2)}({\bm e}_z) = \left( \begin{array}{c}
                                   1 \\[0.4cm]
                                   0
                                   \end{array}
                          \right) \, , \: \: \:
   w^{(-1/2)}({\bm e}_z) = \left( \begin{array}{c}
                                   0 \\[0.4cm]
                                   1
                                   \end{array}
                          \right)
\end{eqnarray}
which describes the spin--up, $\sigma = +1/2$, and spin--down, $\sigma = -1/2$, states along the $z$--axis. With the help of the solutions (\ref{eq_Dirac_spinor}) we can construct now the spinor $w^{(\lambda)}({\bm n})$ for the electron propagating in some arbitrary direction ${\bm n} = (\sin\theta \cos\varphi, \, \sin\theta \sin\varphi, \, \cos\theta)$ with regard to the quantization axis:
\begin{eqnarray}
   \label{eq_Dirac_spinor_arbitrary_direction}
   w^{(\lambda)}({\bm n}) &=& \sum\limits_{\sigma = - 1/2}^{1/2} D^{1/2}_{\sigma \lambda}(\varphi, \theta, 0) w^{(\sigma)}({\bm e}_z) \nonumber \\
   &=& \sum\limits_{\sigma = - 1/2}^{1/2} {\rm e}^{-i \sigma \varphi} d^{1/2}_{\sigma \lambda}(\theta) w^{(\sigma)}({\bm e}_z) \, .
\end{eqnarray}
Here $D^{1/2}_{\sigma \lambda}(\varphi, \theta, 0) = {\rm e}^{- i \sigma \varphi} \, d^{\, 1/2}_{\sigma \lambda}(\theta)$ is the Wigner $D$--function, and $d^{\, 1/2}_{\sigma \lambda}(\theta) = \delta_{\sigma \lambda} \cos\left(\theta/2\right) - 2 \sigma \delta_{\sigma, -\lambda} \sin\left(\theta/2\right)$, see Ref.~\cite{VaM88}.

Making use of the expansion (\ref{eq_Dirac_spinor_arbitrary_direction}) and Eqs.~(\ref{eq_plane_wave_functions})--(\ref{eq_Dirac_bispinor}) we can finally derive the wave--function of the initial free electron state:
\begin{eqnarray}
   \label{eq_plane_wave_initial}
   \psi_{{\bm p} \lambda}({\bm r}) &=& {\rm e}^{i {\bm p} {\bm r}} \,  \sum\limits_{\sigma = - 1/2}^{1/2}  {\rm e}^{-i \sigma \varphi} d^{1/2}_{\sigma \lambda}(\theta) \nonumber \\
   &\times& \left( \begin{array}{c}
                                   \sqrt{\varepsilon + m_e} \, w^{(\sigma)}({\bm e}_z) \\[0.4cm]
                                   2 \lambda \sqrt{\varepsilon - m_e} \, w^{(\sigma)}({\bm e}_z)
                                   \end{array}
                          \right) \, ,
\end{eqnarray}
and similar expression for $\psi_{{\bm p}' \lambda'}({\bm r})$. Beside these functions we have to define also the interaction potential $V({\bm r})$ in order to calculate the scattering amplitude (\ref{eq_scattering_amplitude_general}). In the present study, we consider a spherically--symmetric Yukawa potential:
\begin{equation}
   \label{eq_Yukawa_potential}
   V({\bm r}) = - \frac{Z e^2}{r} \, {\rm e}^{-\mu r} \, ,
\end{equation}
as approximation to the Coulomb field of the nucleus that is screened by atomic electrons, see Section~\ref{sec_computations} for further details. By inserting this potential and the wave--functions $\psi_{{\bm p} \lambda}({\bm r})$ and $\psi_{{\bm p}' \lambda'}({\bm r})$, into Eq.~(\ref{eq_scattering_amplitude_general}) we find:
\begin{eqnarray}
   \label{eq_scattering_amplitude_plane_wave}
   f_{\lambda \lambda'}({\bm p}, {\bm p}') &=& \frac{Z e^2}{q^2 + \mu^2} \, u^\dag_{{\bm p}' \lambda'}\, u_{{\bm p} \lambda} \, \nonumber \\
   &=& \frac{2 Z e^2}{q^2 + \mu^2} \, \left( \varepsilon \delta_{\lambda \lambda'} + m_e \delta_{\lambda, -\lambda'} \right) \nonumber \\
   &\times& \sum\limits_{\sigma = - 1/2}^{1/2} {\rm e}^{i \sigma (\varphi' - \varphi)} \, d^{1/2}_{\sigma \lambda}(\theta) \, d^{1/2}_{\sigma \lambda'}(\theta') \, ,
\end{eqnarray}
where ${\bm q} = {\bm p} - {\bm p}'$ is the momentum transfer, and in the last two lines we used the orthonormality of the Pauli spinors, $w^{(\sigma) \dag}({\bm e}_z) \, w^{(\sigma')}({\bm e}_z) = \delta_{\sigma \sigma'}$. In this expression, moreover, the angles $(\theta, \varphi)$ and $(\theta', \varphi')$ characterize the propagation directions of the incident and scattered electrons with regard to the overall coordinate system.

\subsubsection{Cross section and degree of polarization}
\label{subsubsec_cross_section_plane}

With the help of the amplitude (\ref{eq_scattering_amplitude_plane_wave}) we are ready now to investigate the properties of the plane--wave (relativistic) electrons scattered by the Yukawa potential. We start from the angle--differential cross section which is defined in the standard way as:
\begin{eqnarray}
   \label{eq_cross_section_general_plane_wave_definition}
   \frac{{\rm d}\sigma^{\rm (pl)}_{\lambda \lambda'}}{{\rm d}\Omega'} &=& \frac{2p}{j_z} \, \left| f_{\lambda \lambda'}({\bm p}, {\bm p}') \right|^2 \, ,
\end{eqnarray}
where $j_z = \bar{\psi}_{{\bm p} \lambda}({\bm r}) \gamma_z \psi_{{\bm p} \lambda}({\bm r})$ is the current density and $\gamma_z$ is the Dirac matrix. For the initial--state wavefunction $\psi_{{\bm p} \lambda}({\bm r})$ as given by Eq.~(\ref{eq_plane_wave_initial}), the current density trivially simplifies to $j_z = 2 p$, and we find:
\begin{eqnarray}
   \label{eq_cross_section_general_plane_wave}
   \frac{{\rm d}\sigma^{\rm (pl)}_{\lambda \lambda'}}{{\rm d}\Omega'}(\Theta \,) &=& \left| f_{\lambda \lambda'}({\bm p}, {\bm p}') \right|^2 \nonumber \\
   &=& \frac{4 Z^2 e^4}{\left(\bm q^2 + \mu^2\right)^2} \, \left( \varepsilon^2 \delta_{\lambda \lambda'} + m_e^2 \delta_{\lambda, -\lambda'} \right) \nonumber \\
   &\times& \left| \sum\limits_{\sigma = - 1/2}^{1/2} {\rm e}^{i \sigma (\varphi' - \varphi)} \, d^{1/2}_{\sigma \lambda}(\theta) \, d^{1/2}_{\sigma \lambda'}(\theta') \right|^2 \nonumber \\[0.2cm]
   &=& \frac{2 Z^2 e^4}{\left(q^2 + \mu^2\right)^2} \, \Big[ \varepsilon^2 (1 + \cos\Theta)) \delta_{\lambda \lambda'} \nonumber \\[0.2cm]
   &+& m_e^2 (1 - \cos\Theta)) \delta_{\lambda, -\lambda'} \Big] \, ,
\end{eqnarray}
where $\cos\Theta = {\bm p}{\bm p}'/pp'$. Apart from this \textit{scattering} angle $\Theta$, the cross section depends also on the initial and final--state helicities $\lambda$ and $\lambda'$. If the incident electrons are unpolarized and the spin state of the scattered electrons remains unobserved, one derives from Eq.~(\ref{eq_cross_section_general_plane_wave}):
\begin{eqnarray}
   \label{eq_cross_section_general_plane_wave_unpolarized}
   \frac{{\rm d}\sigma^{\rm (pl)}_0}{{\rm d}\Omega'}(\Theta \,) &=& \frac{1}{2} \sum\limits_{\lambda \lambda'}
   \frac{{\rm d}\sigma_{\lambda \lambda'}}{{\rm d}\Omega}(\Theta \,) \nonumber \\ \nonumber \\[-0.2cm]
   &=& \frac{4 Z^2 e^4 m_e^2}{\left({\bm q}^2 + \mu^2\right)^2} \, \left( 1 + \Delta \right) \, ,
\end{eqnarray}
where the relativistic correction factor $\Delta$ is given by:
\begin{equation}
   \label{eq_Delta_function}
   \Delta = \frac{\varepsilon^2 - m_e^2}{m_e^2} \, \cos^2\left(\Theta/2 \right) \, .
\end{equation}
For $\mu = 0$ this formula reduces to the well--known expression for the Mott scattering by the \textit{Coulomb} potential:
\begin{equation}
   \label{eq_cross_section_Mott_Coulomb_potential}
   \frac{{\rm d}\sigma_{\rm Mott}}{{\rm d}\Omega'}(\theta' \,) =
   \frac{Z^2 e^4}{4 v^2 p^2 \, \sin^4 (\theta'/2)} \, \left( 1 - v^2 \sin^2(\theta'/2) \right) \, .
\end{equation}
Here $v = p/\varepsilon$ and we assumed that the incident electrons propagate along the $z$--axis and, hence, $\theta = 0$ and $\Theta = \theta'$.

The cross section (\ref{eq_cross_section_general_plane_wave_unpolarized}) can be simplified also for the low--energy collisions, when $T = \varepsilon - m_e \ll m_e$. In this case the correction factor $\Delta$ from Eq.~(\ref{eq_Delta_function}) vanishes and the ${\rm d}\sigma^{\rm (pl)}_0/{\rm d}\Omega'$ reads as:
\begin{eqnarray}
   \label{eq_cross_section_general_plane_wave_unpolarized_non_relativistic}
   \frac{{\rm d}\sigma^{\rm (pl)}_0}{{\rm d}\Omega'}(\Theta \,)
   &=& \frac{4 Z^2 e^4 m_e^2}{\left({\bm q}^2 + \mu^2\right)^2} \, ,
\end{eqnarray}
which corresponds to the non--relativistic scattering of spinless particles by an Yukawa potential. This simple formula can not be applied, however, to analyze present electron microscope (scattering) experiments in which the incident electrons have a kinetic energy of up to $T = $~300~keV. For such an energy, the correction factor (\ref{eq_Delta_function}) increases to $\Delta \approx 1.5 \cos^2\left(\Theta/2 \right)$ which implies that both the total Mott cross section and the angular distribution of scattered electrons can be strongly affected by the relativistic effects. This stresses the importance of the fully--relativistic analysis of the scattering of high--energetic plane--wave (as well as twisted) electrons.

The helicity--dependent cross section (\ref{eq_cross_section_general_plane_wave}) can be applied also to analyze the degree of longitudinal polarization of the scattered electrons $P$, which is also often referred to as the mean double helicity. For example, if the incident electron beam is completely longitudinally polarized, $\lambda = 1/2$, we find:
\begin{eqnarray}
   \label{eq_degree_polarization_plane_wave}
   P^{\rm (pl)}(\Theta) &=& \frac{{\rm d}\sigma^{\rm (pl)}_{1/2, \, 1/2} - {\rm d}\sigma^{\rm (pl)}_{1/2, \, -1/2}}{{\rm d}\sigma^{\rm (pl)}_{1/2, \, 1/2} + {\rm d}\sigma^{\rm (pl)}_{1/2, \, -1/2}}
   \nonumber \\[0.2cm]
   &=& \frac{R + \cos\Theta}{1 + R \cos\Theta} \, ,
\end{eqnarray}
where:
\begin{equation}
   \label{eq_R_ratio}
   R = \frac{\varepsilon^2 - m_e^2}{\varepsilon^2 + m_e^2} \, .
\end{equation}
As seen from these expressions, in the ultra--relativistic regime $R \to 1$ and the helicity is conserved during the course of scattering, $P(\Theta) = 1$. In contrast, for slow incident electrons $R \to 0$ and, hence, $P = \cos\Theta$ which corresponds to the projection of the initial electron spin onto the final momentum ${\bm p}'$.

\subsection{Twisted electron scattering}
\label{subsec_twisted_wave}

After having briefly recalled the basic relations used to describe the Mott scattering of plane--wave electrons by the central Yukawa potential, we are ready to consider the \textit{twisted} electron beam. Similar to before, we shall start from the derivation of the scattering amplitude and discuss its main properties.

\subsubsection{``Twisted'' transition amplitude}
\label{subsubsec_amplitude_twisted}

In order to derive the amplitude for the Mott scattering of twisted electrons we have to return to the general expression (\ref{eq_scattering_amplitude_general}). In this formula, the initial--state wave--function $\psi_{{\bm p} \lambda}({\bm r})$ should be modified to represent a twisted state. Here we assume that the incident twisted electrons propagate along the quantization ($z$--) axis and have well--defined values of (i) the longitudinal linear momentum $p_z$, (ii) the modulus of the transverse momentum $|{\bm p}_\perp| = \varkappa$, and (iii) the half--integer projection of the total angular momentum, $J_z = m$. Such a \textit{Bessel} state has, moreover, the definite energy $\varepsilon = \sqrt{\varkappa^2 + p_z^2 + m_e^2}$ and helicity $\lambda$, and is described by the wave--function:
\begin{equation}
   \label{eq_wave_function_twisted_electrons}
   \psi_{\varkappa m p_z \lambda}({\bm r}) = \int{\frac{{\rm d}^2 {\bm p}_\perp}{(2 \pi)^2}} \, a_{\varkappa m}({\bm p}_\perp) \,
   u_{{\bm p} \lambda} \, {\rm e}^{i {\bm p} {\bm r}} \, ,
\end{equation}
see Refs.~\cite{JeS11b,IvS11,MaH14} and Appendix \ref{appendix_wave_function} for further details. As seen from this expression, the $\psi_{\varkappa m p_z \lambda}({\bm r})$ can be considered as a coherent superposition of the plane--waves $u_{{\bm p} \lambda} \, {\rm e}^{i {\bm p} {\bm r}}$, weighted with the amplitude:
\begin{equation}
   \label{eq_a_amplitude}
   a_{\varkappa m}({\bm p}_\perp) = (-i)^m \, {\rm e}^{i m \varphi_p} \, \sqrt{\frac{2 \pi}{\varkappa}} \, \delta\left(|{\bm p}_\perp| - \varkappa \right) \, .
\end{equation}
The linear momenta of these plane--wave components, ${\bm p} = \left( {\bm p}_\perp, p_z \right) = \left( \varkappa \cos\varphi_p, \varkappa \sin\varphi_p, p_z \right)$, form the surface of a cone with the opening angle $\theta_p = \arctan (\varkappa / p_z)$.

Employing the initial--state wave--function (\ref{eq_wave_function_twisted_electrons}) we find the amplitude for the Mott scattering of the Bessel electrons:
\begin{eqnarray}
   \label{eq_scattering_amplitude_twisted_wave}
   F^{(m)}_{\lambda \lambda'}({\bm p}, {\bm p}', {\bm b}) &=& -\frac{1}{4 \pi} \int{\psi_{{\bm p}' \lambda'}^\dag({\bm r})
   \, V({\bm r}) \, \psi_{\varkappa m p_z \lambda}({\bm r}) \, {\rm d}{\bm r}} \nonumber \\[0.2cm]
   && \hspace*{-1.7cm} = \int{\frac{{\rm d}^2 {\bm p}_\perp}{(2 \pi)^2}} \, a_{\varkappa m}({\bm p}_\perp) \, {\rm e}^{- i {\bm p}_\perp {\bm b}}
   f_{\lambda \lambda'}({\bm p}, {\bm p}') \nonumber \\[0.2cm]
   && \hspace*{-1.7cm} = (-i)^m \, \sqrt{\frac{\varkappa}{2 \pi}} \int\limits_{0}^{2\pi} \frac{{\rm d}\varphi_p}{2 \pi} \, {\rm e}^{i m \varphi_p - i {\bm p}_\perp {\bm b}}
   \, f_{\lambda \lambda'}({\bm p}, {\bm p}') \, ,
\end{eqnarray}
where $f_{\lambda \lambda'}({\bm p}, {\bm p}')$ is the ``standard'' plane--wave matrix element (\ref{eq_scattering_amplitude_plane_wave}). In Eq.~(\ref{eq_scattering_amplitude_twisted_wave}), moreover, we have introduced the exponential factor ${\rm exp}(-i {\bm p}_{\perp} {\bm b})$ to specify the lateral position of the scatterer atom with regard to the central ($z$--) axis of the incident electron beam, and where ${\bm b} = (b_x, b_y ,0)$ is the impact parameter. The introduction of this factor reflects the fact that, in contrast to the plane--wave case, the Bessel beam has a complex spatial structure in the plane perpendicular to the $z$--axis \cite{IvS11,MaH14}.

In order to further evaluate the $F^{(m)}_{\lambda \lambda'}({\bm p}, {\bm p}', {\bm b})$ we have to insert the plane--wave amplitude (\ref{eq_scattering_amplitude_plane_wave}) into Eq.~(\ref{eq_scattering_amplitude_twisted_wave}). After performing some simple algebraic manipulations, we finally obtain:
\begin{eqnarray}
   \label{eq_scattering_amplitude_twisted_wave_final}
   F^{(m)}_{\lambda \lambda'}({\bm p}, {\bm p}', {\bm b}) &=& 2 Z e^2 \, i^{-m} \, \sqrt{\frac{\varkappa}{2 \pi}} \,
   \left(\varepsilon \delta_{\lambda \lambda'} + m_e \delta_{\lambda, -\lambda'} \right) \nonumber\\[0.2cm]
   && \hspace*{-2.1cm} \times {\rm e}^{i m \varphi'} \, \sum\limits_{\sigma = - 1/2}^{1/2} \, d^{1/2}_{\sigma \lambda}(\theta_p) \, d^{1/2}_{\sigma \lambda'}(\theta') \, I_{m - \sigma}(\alpha, \beta, {\bm b}) \,  ,
\end{eqnarray}
where the function $I_{m - \sigma}(\alpha, \beta, {\bm b})$ reads as:
\begin{equation}
   \label{eq_function_I_b}
   I_{n}(\alpha, \beta, {\bm b}) = \int\limits_{0}^{2 \pi} \frac{{\rm d}\phi}{2 \pi} \, \frac{{\rm e}^{i\, n \phi - i \varkappa b \cos(\phi + \varphi' - \varphi_b)}}{\alpha - \beta \cos\phi} \, .
\end{equation}
In this expression, $\varphi_b$ is the azimuthal angle of the impact--parameter vector ${\bm b}$, and we have introduced the short--hand notations:
\begin{eqnarray}
   \label{eq_alpha}
   \alpha &=& 2p^2 - 2 p_z p'_z + \mu^2 \nonumber \\
   				&=& 2p^2 (1 - \cos\theta_p \cos\theta') + \mu^2 \, , \\
   \beta &=& 2 \varkappa p'_{\perp} = 2p^2 \sin\theta_p \sin\theta' \, ,
   \label{eq_beta}
\end{eqnarray}
where both $\alpha$ and $\beta$ are positive.

Eqs.~(\ref{eq_scattering_amplitude_twisted_wave_final})--(\ref{eq_function_I_b}) represent the amplitude for the Mott scattering of twisted (Bessel) electrons by the Yukawa potential (\ref{eq_Yukawa_potential}), whose center is shifted by ${\bm b}$ with respect the $z$--axis. In the following Section we will apply this amplitude to study the angular distribution of the scattered electrons.

\subsubsection{Impact--parameter dependent angular distribution}
\label{subsubsec_cross_section_twisted_b_dependence}

In contrast to the plane--wave case (\ref{eq_cross_section_general_plane_wave_definition}), the evaluation of the cross section for the scattering of a vortex beam by a single well--localized atom is not a simple task. It requires the re--definition of the incident flux $j_z$, which is now a function of the impact parameter ${\bm b}$ and, more generally, the \textit{concept} of the cross section. The derivation of such an impact--parameter--dependent Mott cross section is out of scope of the present study and will be discussed separately \cite{KoS15}. Here, instead, we will focus on the angular distribution of the scattered electrons:
\begin{eqnarray}
   \label{eq_angular_distribution_twisted}
   W^{\rm (tw)}_{\lambda, \lambda'}(\theta', \varphi'; \, {\bm b}, m) &=& \mathcal{N} \left| F^{(m)}_{\lambda \lambda'}({\bm p}, {\bm p}', {\bm b}) \right|^2 \nonumber \\
   && \hspace*{-2.5cm} = \mathcal{N} \, \frac{2 Z^2 e^4}{\pi} \, \varkappa \, (\varepsilon^2 \delta_{\lambda \lambda'} + m_e^2 \delta_{\lambda, -\lambda'})
   \nonumber \\
   && \hspace*{-2.5cm} \times \left| \sum\limits_{\sigma = - 1/2}^{1/2} \, d^{1/2}_{\sigma \lambda}(\theta_p) \, d^{1/2}_{\sigma \lambda'}(\theta') \, I_{m - \sigma}(\alpha, \beta, {\bm b}) \right|^2 \, ,
\end{eqnarray}
where the pre--factor $\mathcal{N}$ is defined by the normalization condition $\int{W^{\rm (tw)}_{\lambda, \lambda'}(\theta', \varphi'; \, {\bm b}, m) \, {\rm d}\Omega'} = 1$.

From the discussion above and Eq.~(\ref{eq_angular_distribution_twisted}), it is apparent that the angular distribution $W^{\rm (tw)}_{\lambda, \lambda'}(\theta', \varphi'; \, {\bm b}, m)$ of the outgoing electrons depends on the position of the Yukawa scatterer within the incident wave front. Even though the detailed study of such a ${\bm b}$--dependence requires numerical computations and will be presented in Section~\ref{sec_results}, here we consider the special case of a central collision, ${\bm b} = 0$. For this scenario, one can evaluate the scattering amplitude (\ref{eq_scattering_amplitude_twisted_wave_final}) and, hence, the $W^{\rm (tw)}_{\lambda, \lambda'}(\theta', \varphi'; \, {\bm b}, m)$ analytically. In particular, we re--write the integral (\ref{eq_function_I_b}), which enters in $F^{(m)}_{\lambda \lambda'}({\bm p}, {\bm p}', {\bm b})$, as:
\begin{eqnarray}
   \label{eq_function_I_b_0}
   I_{n}(\alpha, \beta, {\bm b} = 0) &=& \int\limits_{0}^{2 \pi} \frac{{\rm d}\phi}{2 \pi} \, \frac{{\rm e}^{i\, n \phi}}{\alpha - \beta \cos\phi}
   \nonumber \\
   &=& \frac{1}{i \pi \beta} \oint \frac{z^{|n|}}{(z_1 - z)(z - z_2)} \, {\rm d}z \, ,
\end{eqnarray}
where in the second line we introduced the complex variable $z = {\rm e}^{i \phi}$. The integration contour in Eq.~(\ref{eq_function_I_b_0}) is the unit circle and the positions of the poles are given by $z_{1,2} = \left(\alpha \pm \sqrt{\alpha^2 - \beta^2} \right)/\beta$ with $z_1 > 1$ and $z_2 < 1$. By noting that the integrand in Eq.~(\ref{eq_function_I_b_0}) has a single simple pole inside the unit circle, and taking its residue at $z_2$, we find:
\begin{equation}
   \label{eq_function_I_b_0_final}
   I_{n}(\alpha, \beta, {\bm b} = 0) = \frac{1}{\sqrt{\alpha^2 - \beta^2}} \, \left( \frac{\beta}{\alpha + \sqrt{\alpha^2 - \beta^2}} \right)^{|n|}
   \, .
\end{equation}
With the help of this expression one can easily evaluate the angular distribution (\ref{eq_angular_distribution_twisted}) of the scattered electrons for \textit{zero} impact parameter ${\bm b} = 0$. Of special interest here is the behaviour of $W^{\rm (tw)}_{\lambda, \lambda'}(\theta', \varphi'; \, {\bm b} = 0, m)$ at small scattering angles with respect to the incident beam axis. If we perform the Taylor expansion of
Eq.~(\ref{eq_angular_distribution_twisted}) for $\theta' \to 0$ we obtain:
\begin{equation}
   \label{eq_angular_distribution_twisted_b_0_asymptotic}
   W^{\rm (tw)}_{\lambda, \lambda'}(\theta', \varphi'; \, {\bm b} = 0, m) \propto \left( \theta' \right)^{2 |m - \lambda'|} ,
\end{equation}
which indicates that the angular distribution of the scattered electrons vanishes for the forward emission if the incident beam is twisted, ${\bm b} = 0$ and $m \ne \lambda'$. This behaviour has been predicted recently also within the non--relativistic framework \cite{VaB13}, and is strongly different from what is expected for the ``plane--wave'' angle--differential cross section (\ref{eq_cross_section_general_plane_wave}) which is maximal for $\theta' = 0$.

The dip (\ref{eq_angular_distribution_twisted_b_0_asymptotic}) in the electron angular distribution for $\theta' = 0$ disappears with the increase of the impact parameter. Indeed, by making use of Eq.~(\ref{eq_scattering_amplitude_twisted_wave_final}) we find that the angular distribution:
\begin{eqnarray}
   \label{eq_angular_distribution_twisted_bne0}
   W^{\rm (tw)}_{\lambda, \lambda'}(\theta' = 0, \varphi'; \, {\bm b}, m) &=& \mathcal{N} \, \frac{1}{\pi}
   \frac{Z^2e^4 \varkappa\,\left[ J_{m-\lambda'}(\varkappa b)\right]^2}{(4p^2\sin^2{(\theta_p/2)}+\mu^2)^2} \nonumber \\[0.2cm]
   && \hspace*{-3.8cm} \times \left[\varepsilon^2(1+\cos\theta_p)\delta_{\lambda\lambda'}
   + m_e^2(1-\cos\theta_p)\delta_{\lambda,-\lambda'}\right] \, ,
\end{eqnarray}
does not vanish for the forward scattering and $b \ne 0$. For small impact parameters, $b \ll 1/\varkappa$, this expression predicts that the electron emission quickly increases with the impact parameter, $W^{\rm (tw)}_{\lambda, \lambda'}(\theta' = 0, \varphi'; \, {\bm b}, m) \propto b^{2|m - \lambda'|}$.

\subsubsection{Averaging over the impact parameter}
\label{subsubsec_cross_section_twisted_b_averaged}

Until now, we have discussed the evaluation of the angular distribution of scattered electrons for the case when the incident twisted beam collides with a \textit{single} and well localized Yukawa potential. This scenario, however, can be hardly realized in nowadays Mott scattering experiments in which thin foils are usually used as a target. We can describe the (solid--state) target by an ensemble of Yukawa scatterers that are randomly and uniformly distributed over the transverse extent of the incident beam. If one neglects collective and multiple scattering effects, the differential cross section for the Mott scattering of twisted electrons by such a ``foil'' can be introduced in a way similar to that of the standard plane--wave case (\ref{eq_cross_section_general_plane_wave_definition}). That is, by averaging the (square of the) transition amplitude (\ref{eq_scattering_amplitude_twisted_wave}) over the impact parameter ${\bm b}$ we obtain:
\begin{equation}
   \label{eq_cross_section_twisted_averaged_general}
   \frac{{\rm d}\sigma^{\rm (tw)}_{\lambda \lambda'}}{{\rm d}\Omega'}(\theta' , \, \theta_p) 
   		= \frac{2p}{\mathcal J_z} \int{\left| F^{(m)}_{\lambda \lambda'}({\bm p}, {\bm p}', {\bm b}) \right|^2
   		 {\rm d}^2{\bm b}} \, ,
\end{equation}
where $\mathcal J_z$ is the total incident current:
\begin{eqnarray}
   \label{eq_total_incident_current}
  \mathcal  J_z &=& \int j_z({\bm b}) {\rm d}^2{\bm b} \nonumber \\
   &\equiv& \int \bar{\psi}_{\varkappa m p_z \lambda}({\bm r}-{\bm b}) \, \gamma_z \, \psi_{\varkappa m p_z \lambda}({\bm r}-{\bm b}) {\rm d}^2{\bm b} \, ,
\end{eqnarray}
and $\psi_{\varkappa m p_z \lambda}({\bm r}-{\bm b})$ is the wave--function of the electron vortex beam whose axis is shifted by ${\bm b}$ with respect to the overall $z$--axis.

In order to further evaluate the Mott scattering cross section we should perform the integration over the impact parameter ${\bm b}$ in Eqs.~(\ref{eq_cross_section_twisted_averaged_general}) and (\ref{eq_total_incident_current}). Assuming that the radius $R$ describes the characteristic transverse size of the Bessel beam and $R \gg 1/\varkappa$, the total current $\mathcal J_z$ can be approximated as:
\begin{eqnarray}
   \label{eq_total_incident_current_approximation}
   \mathcal J_z &\simeq& \int \bar{\psi}_{\varkappa m p_z \lambda}(-{\bm b}) \, \gamma_z \, \psi_{\varkappa m p_z \lambda}(-{\bm b}) {\rm d}^2{\bm b} \nonumber \\[0.2cm]
   &=& 2p_z \int \left| a_{\varkappa m}({\bm p}_{\perp}) \right|^2 \, \frac{{\rm d}{\bm p}_{\perp}}{(2 \pi)^2} = 2 p_z \frac{R}{\pi} \, ,
\end{eqnarray}
where in the second line we have employed Eq.~(76a) from \cite{JeS11b} to evaluate the square of the $\delta$ function,
\begin{equation}
   \label{eq_delta_squared}
   \left| \delta(|{\bm p}_\perp| - \varkappa) \right|^2 = \frac{R}{\pi} \, \delta(|{\bm p}_\perp| - \varkappa) \, .
\end{equation}
By making use of this expression and of Eq.~(\ref{eq_scattering_amplitude_twisted_wave}) one can also compute the integral of the square of the transition amplitude:
\begin{eqnarray}
   \label{eq_integral_square_amplitude}
   \int{\left| F^{(m)}_{\lambda \lambda'}({\bm p}, {\bm p}', {\bm b}) \right|^2 {\rm d}^2{\bm b}} && \nonumber\\[0.2cm]
   && \hspace*{-3cm} = \frac{R}{\pi}
   \, \int\limits_{0}^{2\pi} \left|f_{\lambda \lambda'}({\bm p}, {\bm p}') \right|^2 \, \frac{{\rm d}\varphi_p}{2 \pi} \, ,
\end{eqnarray}
and, hence, the angle--differential cross section:
\begin{eqnarray}
   \label{eq_cross_section_twisted_averaged_1}
   \frac{{\rm d}\sigma^{\rm (tw)}_{\lambda \lambda'}}{{\rm d}\Omega'}(\theta' , \, \theta_p) = \frac{1}{\cos\theta_p}
   \, \int\limits_{0}^{2\pi} \left|f_{\lambda \lambda'}({\bm p}, {\bm p}') \right|^2 \, \frac{{\rm d}\varphi_p}{2 \pi} \, .
\end{eqnarray}
To carry out the remaining integration over $\varphi_p$ we insert here the explicit form of the plane--wave transition amplitude (\ref{eq_scattering_amplitude_plane_wave}):
\begin{eqnarray}
   \label{eq_cross_section_twisted_averaged_2}
   \frac{{\rm d}\sigma^{\rm (tw)}_{\lambda \lambda'}}{{\rm d}\Omega'}(\theta' , \, \theta_p) &=& \frac{4 Z^2 e^4}{\cos\theta_p} \,
   \left( \varepsilon^2 \delta_{\lambda \lambda'} + m_e^2 \delta_{\lambda, -\lambda'} \right) \nonumber \\
   &\times& \sum\limits_{\sigma \sigma'} d^{1/2}_{\sigma \lambda}(\theta_p) \, d^{1/2}_{\sigma \lambda'}(\theta')
   \, d^{1/2}_{\sigma' \lambda}(\theta_p) \, d^{1/2}_{\sigma' \lambda'}(\theta') \nonumber \\
   &\times&  \int\limits_{0}^{2\pi} \frac{{\rm e}^{i(\sigma - \sigma')(\varphi_p - \varphi')}}{({\bm q}^2 + \mu^2)^2} \frac{{\rm d}\varphi_p}{2 \pi} \, ,
\end{eqnarray}
and observe that the integral in the last line of this expression is given by:
\begin{eqnarray}
   \label{eq_integral_2}
   \int\limits_{0}^{2\pi} \frac{{\rm e}^{i(\sigma - \sigma')(\varphi_p - \varphi')}}{({\bm q}^2 + \mu^2)^2} \frac{{\rm d}\varphi_p}{2 \pi} &=&
   -\frac{\partial}{\partial \alpha} I_{\sigma' - \sigma}(\alpha, \beta, 0) \nonumber \\
   &=& \frac{\alpha \, \delta_{\sigma \sigma'} + \beta \, \delta_{\sigma, -\sigma'}}{(\alpha^2 - \beta^2)^{3/2}} \, ,
\end{eqnarray}
where the function $I_{\sigma' - \sigma}(\alpha, \beta, 0)$ and coefficients $\alpha$ and $\beta$ are defined by Eqs.~(\ref{eq_function_I_b}), (\ref{eq_alpha}) and (\ref{eq_beta}), respectively.

By inserting the integral (\ref{eq_integral_2}) into Eq.~(\ref{eq_cross_section_twisted_averaged_2}) we derive the final expression for the angle--differential cross section of the Mott scattering of the twisted electrons by a macroscopic target:
\begin{eqnarray}
   \label{eq_cross_section_twisted_averaged_final}
   \frac{{\rm d}\sigma^{\rm (tw)}_{\lambda \lambda'}}{{\rm d}\Omega'}(\theta' , \, \theta_p) &=& \frac{4 Z^2 e^4}{\cos\theta_p} \,
   \left( \varepsilon^2 \delta_{\lambda \lambda'} + m_e^2 \delta_{\lambda, -\lambda'} \right) \nonumber \\
   &\times& \sum\limits_{\sigma \sigma'} d^{1/2}_{\sigma \lambda}(\theta_p) \, d^{1/2}_{\sigma \lambda'}(\theta')
   \, d^{1/2}_{\sigma' \lambda}(\theta_p) \, d^{1/2}_{\sigma' \lambda'}(\theta') \nonumber \\
   &\times& \, \frac{\alpha \, \delta_{\sigma \sigma'} + \beta \, \delta_{\sigma, -\sigma'}}{(\alpha^2 - \beta^2)^{3/2}} \, ,
\end{eqnarray}
where the polar angle $\theta'$ of the outgoing electrons is defined with regard to the $z$--axis. As seen from Eq.~(\ref{eq_cross_section_twisted_averaged_final}), the ${\rm d}\sigma^{\rm (tw)}_{\lambda \lambda'}/{\rm d}\Omega'$ is insensitive to the TAM projection $m$ of the (initial) Bessel beam but depends on its opening angle $\theta_p$. If $\theta_p = 0$ and, hence, $\alpha = 4 p^2 \sin^2(\theta'/2) + \mu^2$, $\beta = 0$, and $d^{1/2}_{\sigma \lambda}(\theta_p) = \delta_{\sigma \lambda}$, this cross section expectedly coincides with the plane--wave result (\ref{eq_cross_section_general_plane_wave}) where $\Theta = \theta'$.

The angle--differential cross section (\ref{eq_cross_section_twisted_averaged_final}) still depends on the helicities of the initial-- and final--state electrons, $\lambda$ and $\lambda'$. It can be used, therefore, to evaluate the angular and polarization properties of the scattered electrons for every possible experimental setup. In Section~\ref{sec_results}, for example, we will discuss the angular distribution of the scattered electrons under the assumption that their spin state remains unobserved. For this case, the Mott cross section:
\begin{eqnarray}
   \label{eq_cross_section_twisted_averaged_unpolarizaed_scattered}
   \frac{{\rm d}\sigma^{\rm (tw)}_{\lambda}}{{\rm d}\Omega'}(\theta' , \, \theta_p) &=& \sum\limits_{\lambda'} \frac{{\rm d}\sigma^{\rm tw}_{\lambda \lambda'}}{{\rm d}\Omega}(\theta' , \, \theta_p) \\
   && \hspace*{-2.5cm} = \frac{2 Z^2 e^4}{(\alpha^2 - \beta^2)^{3/2} \, \cos\theta_p} \,
   \Big[\left(\varepsilon^2 + m_e^2 \right) \alpha \nonumber \\
   && \hspace*{-2.5cm} + p^2 \left( \alpha \cos(\theta_p/2) \cos(\theta'/2) + \beta \sin(\theta_p/2) \sin(\theta'/2) \right) \Big] \,  \nonumber
\end{eqnarray}
is insensitive to the helicity of the incoming beam and, within the limit $p^2 \ll m_e^2$, reduces to the non--relativistic expression:
\begin{eqnarray}
   \label{eq_cross_section_twisted_averaged_unpolarizaed_non-relativistic}
   \frac{{\rm d}\sigma^{\rm (tw)}_{\lambda}}{{\rm d}\Omega'}(\theta' , \, \theta_p) &=& \frac{4 Z^2 e^4}{\cos\theta_p} \, \frac{\alpha}{(\alpha^2 - \beta^2)^{3/2}} \, ,
\end{eqnarray}
which describes the scattering of spinless particles by the Yukawa potential \cite{KoS15}.

Similar to the plane--wave case, we can employ Eq.~(\ref{eq_cross_section_twisted_averaged_final}) also to evaluate the degree of longitudinal polarization of the outgoing electrons. This degree $P^{\rm (tw)}(\theta', \theta_p)$ is given by the first line of Eq.~(\ref{eq_degree_polarization_plane_wave}) where the plane--wave cross sections are substituted by ${\rm d}\sigma^{\rm (tw)}_{\lambda \lambda'}/{\rm d}\Omega'$. For the sake of shortness, we will not present here the final expression for the $P^{\rm (tw)}(\theta', \theta_p)$ but discuss in Section~\ref{sec_results} the numerical predictions for the polarization of the final--state electrons.

\begin{figure*}[t]
  \includegraphics[width=0.85\linewidth]{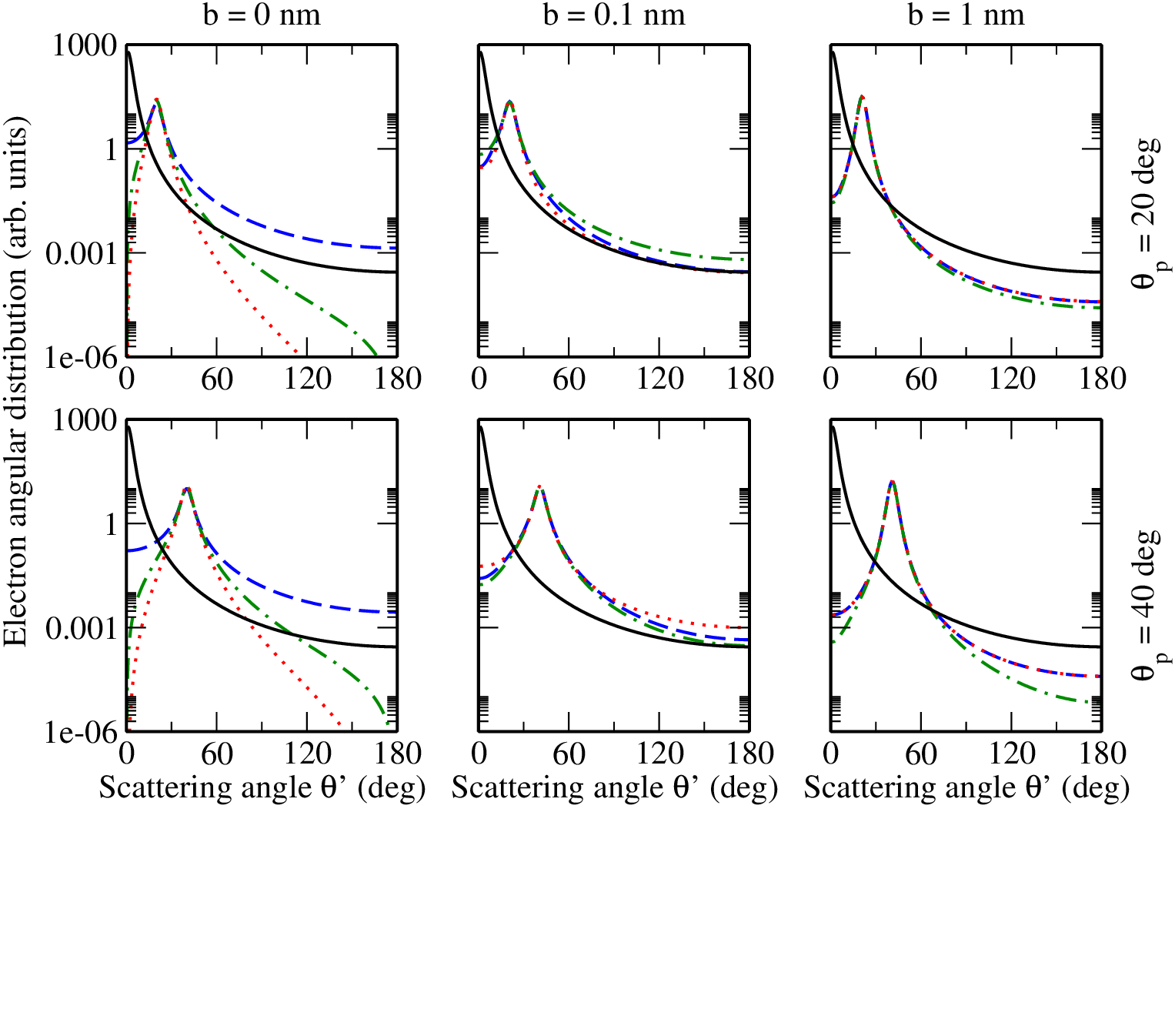}
  \vspace*{-3cm}
  \caption{\label{Fig1}(Color online) The angular distribution (\ref{eq_angular_distribution_twisted_unpolarized}) of the electrons scattered by a \textit{single} hydrogen atom. Calculations have been performed for the incident Bessel electron beam with the kinetic energy 10 keV, helicity $\lambda$~=~1/2, opening angle $\theta_p = 20$~deg (top panels) and $\theta_p = 40$~deg (bottom panels), and the projection of the TAM $m = 1/2$ (blue dashed line), $m = 3/2$ (green dash--dotted line) and $m = 5/2$ (red dotted line). Moreover, the target atom is assumed to be displaced by distances $b$~=~0~nm, $b$~=~0.01~nm, and $b$~=~1~nm with respect to the central beam axis and under the azimuthal angle $\varphi_b = \varphi' = 0$~deg. Results of the calculations are compared with the prediction obtained for the incident plane--wave electrons (solid black curve).}
\end{figure*}
%
%

%
%
%------------------------------------------------------  Computations  ---------------------------------------------------------------
%
%
\section{Scattering by atomic potentials}
\label{sec_computations}

Up to the present we have discussed the Mott scattering of plane--wave and twisted electrons by the Yukawa potential $V(r)$ given by Eq.~(\ref{eq_Yukawa_potential}). In atomic collision studies, the $V(r)$ is often used to approximate (realistic) electron--atom interactions. For example, the electrostatic potential of a neutral atom can be written as a sum of \textit{three} Yukawa terms:
\begin{equation}
   \label{eq_atomic_potential}
   U_{\rm at}(r) = - \frac{Z e^2}{r} \, \sum\limits_{i=1}^3 A_i  \, {\rm e}^{-\mu_i r} \, ,
\end{equation}
where the parameters $A_i$ and $\mu_i$ are determined by a fitting to the results of the Dirac--Hartree--Fock--Slater (DHFS) self--consistent calculations \cite{Sal91,SaM87}. It is rather straightforward to generalize all the results obtained in the previous Section to the case of such an ``atomic'' potential. Indeed, the scattering amplitudes can be written in this case as:
\begin{eqnarray}
   \label{eq_scattering_amplitude_plane_wave_atomic_potential}
   f_{\lambda \lambda'}({\bm p}, {\bm p}') &=& 2 Z e^2 \, \left( \sum\limits_{i = 1}^{3}\frac{A_i}{q^2 + \mu_i^2} \right)
   \, \left( \varepsilon \delta_{\lambda \lambda'} + m_e \delta_{\lambda -\lambda'} \right) \nonumber \\
   &\times& \sum\limits_{\sigma = - 1/2}^{1/2} {\rm e}^{i \sigma (\varphi' - \varphi)} \, d^{1/2}_{\sigma \lambda}(\theta) \, d^{1/2}_{\sigma \lambda'}(\theta') \, ,
\end{eqnarray}
for the plane--wave, and:
\begin{eqnarray}
   \label{eq_scattering_amplitude_twisted_wave_atomic_potential}
   F^{(m)}_{\lambda \lambda'}({\bm p}, {\bm p}', {\bm b}) &=& 2 Z e^2 \, i^{-m} \, \sqrt{\frac{\varkappa}{2 \pi}} \,
   \left(\varepsilon \delta_{\lambda \lambda'} + m_e \delta_{\lambda, -\lambda'} \right) \nonumber\\[0.2cm]
   && \hspace*{-0.0cm} \times {\rm e}^{i m \varphi'} \, \sum\limits_{\sigma = - 1/2}^{1/2} \, d^{1/2}_{\sigma \lambda}(\theta_p) \, d^{1/2}_{\sigma \lambda'}(\theta') \nonumber \\
   && \hspace*{-0.0cm} \times \sum\limits_{i=1}^3 A_i I_{m - \sigma}(\alpha_i, \beta, {\bm b}) \,  ,
\end{eqnarray}
for the twisted--wave electrons, where $\alpha_i = 2p^2 (1 - \cos\theta_p \cos\theta') + \mu_i^2$. With the help of these amplitudes one can again evaluate \textit{analytically} the angular distribution and the (degree of) longitudinal polarization of scattered electrons both for the plane-- and twisted--wave cases. For example, by inserting the expression (\ref{eq_scattering_amplitude_plane_wave_atomic_potential}) into Eq.~(\ref{eq_cross_section_twisted_averaged_1}) we find the angle--differential cross section for the scattering of Bessel electron beam on a macroscopic atomic target:
\begin{eqnarray}
   \label{eq_cross_section_twisted_averaged_atomic}
   \frac{{\rm d}\sigma^{\rm (tw)}_{\lambda \lambda'}}{{\rm d}\Omega'}(\theta' , \, \theta_p) &=& \frac{4 Z^2 e^4}{\cos\theta_p} \,
   \left( \varepsilon^2 \delta_{\lambda \lambda'} + m_e^2 \delta_{\lambda, -\lambda'} \right) \nonumber \\[0.2cm]
   &\times& \sum\limits_{\sigma \sigma'} d^{1/2}_{\sigma \lambda}(\theta_p) \, d^{1/2}_{\sigma \lambda'}(\theta')
   \, d^{1/2}_{\sigma' \lambda}(\theta_p) \, d^{1/2}_{\sigma' \lambda'}(\theta') \nonumber \\
   &\times&  \sum\limits_{i,k =1}^3 A_i \, A_k \, \mathcal{I}_{\sigma' - \sigma}(\alpha_i, \alpha_k, \beta) \, ,
\end{eqnarray}
where the integral $\mathcal{I}_{\sigma' - \sigma}(\alpha_i, \alpha_k, \beta)$ reads as:
\begin{eqnarray}
   \label{eq_integral_atomic}
   \mathcal{I}_{n}(\alpha_i, \alpha_k, \beta) &=& \int\limits_{0}^{2\pi} \frac{{\rm e}^{i n \phi}}{(\alpha_i - \beta \cos\phi) (\alpha_k - \beta \cos\phi)} \frac{{\rm d}\phi}{2 \pi} \nonumber \\
   && \hspace*{-2cm} = \frac{1}{\alpha_k - \alpha_i} \, \int\limits_{0}^{2\pi} \left(\frac{{\rm e}^{i n \phi}}{\alpha_i - \beta \cos\phi} -  \frac{{\rm e}^{i n \phi}}{\alpha_k - \beta \cos\phi}\right) \frac{{\rm d}\phi}{2 \pi} \nonumber \\
   && \hspace*{-2cm} = \frac{I_n(\alpha_i, \beta, 0) - I_n(\alpha_k, \beta, 0)}{\alpha_k - \alpha_i} \, ,
\end{eqnarray}
and $I_n(\alpha, \beta, 0)$ is given by Eq.~(\ref{eq_function_I_b_0_final}).

%
%
%------------------------------------------------------  Results and discussion  ---------------------------------------------------------------
%
%
\section{Results and discussion}
\label{sec_results}

In the previous Sections we have derived the Mott cross section and angular distribution of the outgoing electrons for two cases in which the incident Bessel beam collides with either (i) a single atom, or (ii) a macroscopic target consisting of randomly distributed atoms. Even though the second scenario can be more easily realized experimentally and, hence, is of definite practical interest, let us first discuss the interaction of the twisted wave with a well--localized (single) atom. In Fig.~\ref{Fig1}, for example, we display the angular distribution of electrons scattered by the neutral hydrogen atom and observed by a polarization--insensitive detector:
\begin{equation}
   \label{eq_angular_distribution_twisted_unpolarized}
   W^{\rm (tw)}_{\lambda}(\theta'; \, {\bm b}, m) = \sum\limits_{\lambda'}  W^{\rm (tw)}_{\lambda, \lambda'}(\theta', \varphi' = 0; \, {\bm b}, m) \, .
\end{equation}
Here, $W^{\rm (tw)}_{\lambda, \lambda'}(\theta', \varphi'; \, {\bm b}, m) = {\mathcal N} \left| F^{(m)}_{\lambda \lambda'}({\bm p}, {\bm p}', {\bm b}) \right|^2$ and the amplitude $F^{(m)}_{\lambda \lambda'}$ is given by Eq.~(\ref{eq_scattering_amplitude_twisted_wave_atomic_potential}) with the parameters $\mu_i$ and $A_i$ listed in Table~\ref{Table1}. Calculations have been performed for the incident Bessel beam (\ref{eq_wave_function_twisted_electrons}) with the energy 10~keV, helicity $\lambda$~=~1/2, opening angles $\theta_p$~=~20~deg (top panels) and $\theta_p$~=~40~deg (bottom panels), and three projections of the total angular momentum: $m$~=~1/2 (dashed line), $m$~=~3/2 (dash--dotted line) and $m$~=~5/2 (dotted line). We assume that the hydrogen atom is placed at the azimuthal angle $\varphi_b = \varphi' = 0$~deg and distances $b$~=~0, 0.01 and 1.0 nanometers with respect to the beam axis. As seen from the figure, the angular distribution (\ref{eq_angular_distribution_twisted_unpolarized}) is very sensitive to the parameters of the incident twisted wave as well as to the position of the target atom. For $b = 0$, for example, the forward electron scattering is possible only if the TAM projection of the incident beam is $m$~=~1/2. For $m$~=~3/2 and 5/2, in contrast, the $W^{\rm (tw)}_{\lambda}(\theta'; \, {\bm b}, m)$ vanishes identically if $\theta' \to 0$ as expected from Eq.~(\ref{eq_angular_distribution_twisted_b_0_asymptotic}). If the target atom is shifted from the center of the incident beam, the
angular distribution (\ref{eq_angular_distribution_twisted_unpolarized}) exhibits qualitatively similar $\theta'$--behaviour for all values of $m$. While a forward scattering becomes allowed for $b > 0$, c.f. Eq.~(\ref{eq_angular_distribution_twisted_bne0}), most electrons are emitted under the polar angle $\theta' \simeq \theta_p$ where the $\theta_p$ is the opening angle of the incident twisted beam. One understands such a $\theta'$--behaviour of the angular distribution by noting that the (initial) Bessel state can be \textit{seen} as a coherent superposition of plane waves lying on a momentum cone surface with the opening angle $\theta_p$, see Eq.~(\ref{eq_wave_function_twisted_electrons}). The scattering of a plane--wave component ${\rm e}^{i {\bm p} {\bm r}}$ by the atom results in the emission of the outgoing electron along the linear momentum ${\bm p}$, which is tilted by the angle $\theta_p$ with respect to the overall $z$--axis.

\begin{table}
\begin{center}
\begin{tabular}{c|c|c|c|c|c}
  \hline
  \hline
   &  &  &  &  &  \\[-0.2cm]
  Target & $A_1$ & $\mu_1$ & $A_2$ & $\mu_2$ &  $\mu_3$ \\
  \hline
   &  &  &  &  &  \\[-0.2cm]
  H (Z=1) & -184.39 & 2.0027 & 185.39 & 1.997 & 0 \\[0.2cm]
  Fe (Z=26) & 0.0512 & 31.825 & 0.6995 & 3.7716  & 1.1606 \\[0.2cm]
  \hline
  \hline
\end{tabular}
\caption{Parameters of the effective potential (\ref{eq_atomic_potential}) used in the present calculations. The parameter $A_3 = 1 - A_1 - A_2$. Results are from Ref.~\cite{SaM87}.}
\label{Table1}
\end{center}
\end{table}

\begin{figure}[t]
  \includegraphics[width=0.9\linewidth]{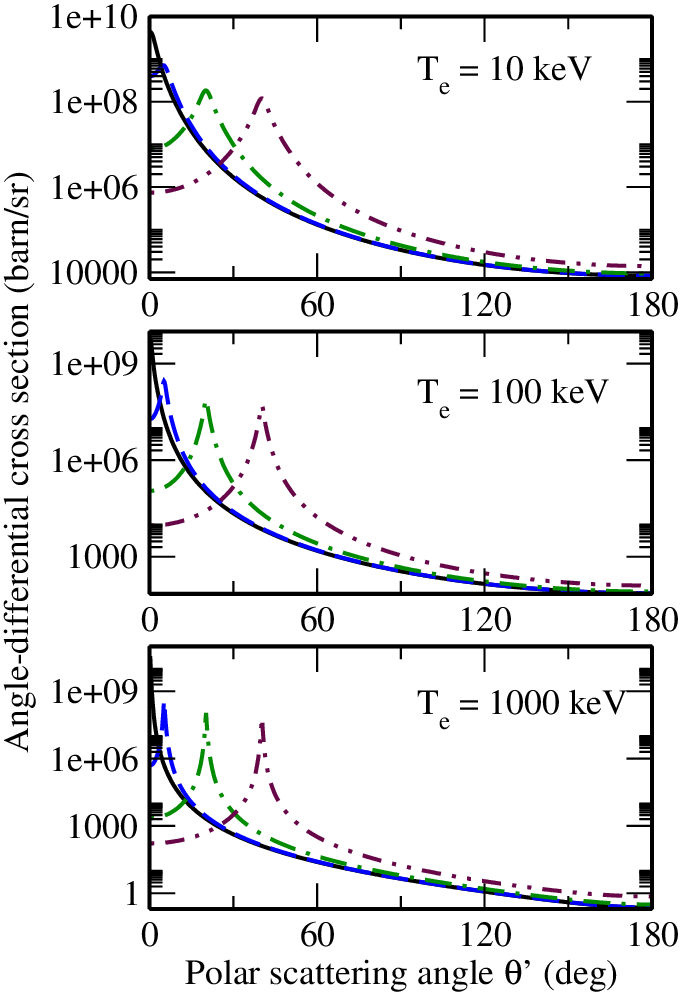}
  \caption{\label{Fig2}(Color online) The angle--differential cross section (\ref{eq_cross_section_twisted_averaged_atomic_unpolarized}) for the Mott scattering of electrons by the macroscopic iron target. Computations have been performed for the incident Bessel beam with the helicity $\lambda$~=~1/2, energies $T_e = 10$~keV (upper panel), 100~keV (middle panel) and 1000~keV (bottom panel), and the opening angles $\theta_p = 5$~deg (blue dashed line), 20~deg (green dash--dotted line) and 40~deg (maroon dash--double--dotted line). Results of the calculations are compared with the prediction obtained for the incident plane--wave electrons (solid black curve).}
\end{figure}

Up to the present, we have discussed the calculations for the scattering of the twisted electrons by the single well--localized hydrogen atom. In more experimentally realistic scenario, the incident Bessel beam collides with a macroscopic target. By making use of Eq.~(\ref{eq_cross_section_twisted_averaged_atomic}) and parameters given in Table~\ref{Table1} one can investigate the angle--differential Mott cross section for this (realistic) case. For example, Fig.~\ref{Fig2} displays the results obtained for the iron target and kinetic electron energies in the range 10~$\le T_e \le$~1000~keV. Similar to before, we assumed that the polarization state of the scattered electrons remains unobserved and, hence, summed over the helicity $\lambda'$:
\begin{equation}
   \label{eq_cross_section_twisted_averaged_atomic_unpolarized}
   \frac{{\rm d}\sigma^{\rm (tw)}_{\lambda}}{{\rm d}\Omega'}(\theta', \theta_p) = \sum\limits_{\lambda'} \frac{{\rm d}\sigma^{\rm (tw)}_{\lambda \lambda'}}{{\rm d}\Omega'}(\theta', \theta_p) \, .
\end{equation}
As mentioned already in Section~\ref{sec_theory}, the averaging over the impact parameters of the target atoms leads to the fact that this cross section ${\rm d}\sigma^{\rm (tw)}_{\lambda}/{\rm d}\Omega'$ is insensitive to the projection $m$ of the TAM and depends only on the opening angle of the incident Bessel beam $\theta_p$. To explore the $\theta_p$--dependence, here we present results for three opening angles: $\theta_p = $~5~deg (dashed line), 20~deg (dash--dotted line) and 40~deg (dotted line). As seen from the figure, the ${\rm d}\sigma^{\rm (tw)}_{\lambda}/{\rm d}\Omega'$ is again peaked at the angle $\theta' = \theta_p$ and the effect becomes more pronounced with the increase of the incoming electron energy. This is in contrast to calculations for the incident plane--wave beam which predict the predominant forward emission, c.f. solid line in Fig.~\ref{Fig2}. Therefore, the most remarkable distinction between the angle--differential cross sections for the scattering of twisted and plane--wave electrons can be observed for large values of the angle $\theta_p$.

\begin{figure}[t]
  \includegraphics[width=0.9\linewidth]{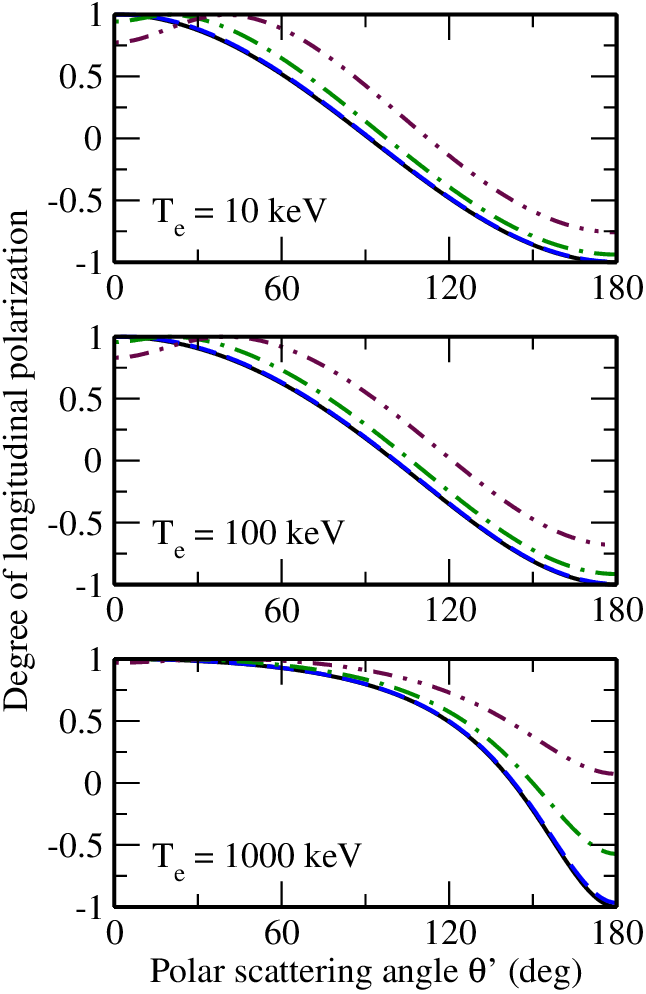}
  \caption{\label{Fig3}(Color online) The degree of longitudinal polarization of electrons scattered by the macroscopic iron target. Calculations have been performed for the same set of parameters as in Fig.~\ref{Fig2}.}
\end{figure}

We can use Eq.~(\ref{eq_cross_section_twisted_averaged_atomic}) to explore not only the angle--differential Mott cross section but also the degree of longitudinal polarization of scattered electrons. If we insert ${\rm d}\sigma^{\rm (tw)}_{\lambda \lambda'}/{\rm d}\Omega'(\theta'; \theta_p)$ into Eq.~(\ref{eq_degree_polarization_plane_wave}) we find the polarization of scattered electrons $P^{\rm (tw)}(\theta; \theta_p)$ for the interaction of the Bessel beam with a macroscopic atomic target. We have computed this degree of polarization again for the iron target and for the same beam parameters as in Fig.~\ref{Fig2}. Results of our calculations are presented in Fig.~\ref{Fig3} and indicate that the $P^{\rm (tw)}(\theta; \theta_p)$ can be rather sensitive to the opening angle $\theta_p$ of the Bessel beam. In particular, while for the scattering of the plane--wave electrons with helicity $\lambda = 1/2$ the polarization decrease monotonically from $P^{\rm (pl)} = 1$ for $\theta' = 0$~deg to $P^{\rm (pl)} = -1$ for $\theta' = 180$~deg, the $P^{\rm (tw)}(\theta; \theta_p)$ is maximal at $\theta' = \theta_p$ and falls down for the forward and backward emission. The influence of the opening angle $\theta_p$ on the longitudinal polarization of outgoing electrons can be observed most easily for low collision energies \textit{and} relatively small scattering angles, $\theta' \lesssim 60$~deg. In this parameter range both the Mott cross section and the variation of $P^{\rm (tw)}(\theta; \theta_p)$ with $\theta_p$ are large. For example, for $T_e =$~10~keV and the observation angle $\theta' =$~40~deg, the degree of polarization increases from 76~\% to almost 100~\% if instead of the plane--wave-- the twisted--wave electrons with the opening angle $\theta_p =$~40~deg collide with the target.

\begin{figure}[t]
  \includegraphics[width=0.9\linewidth]{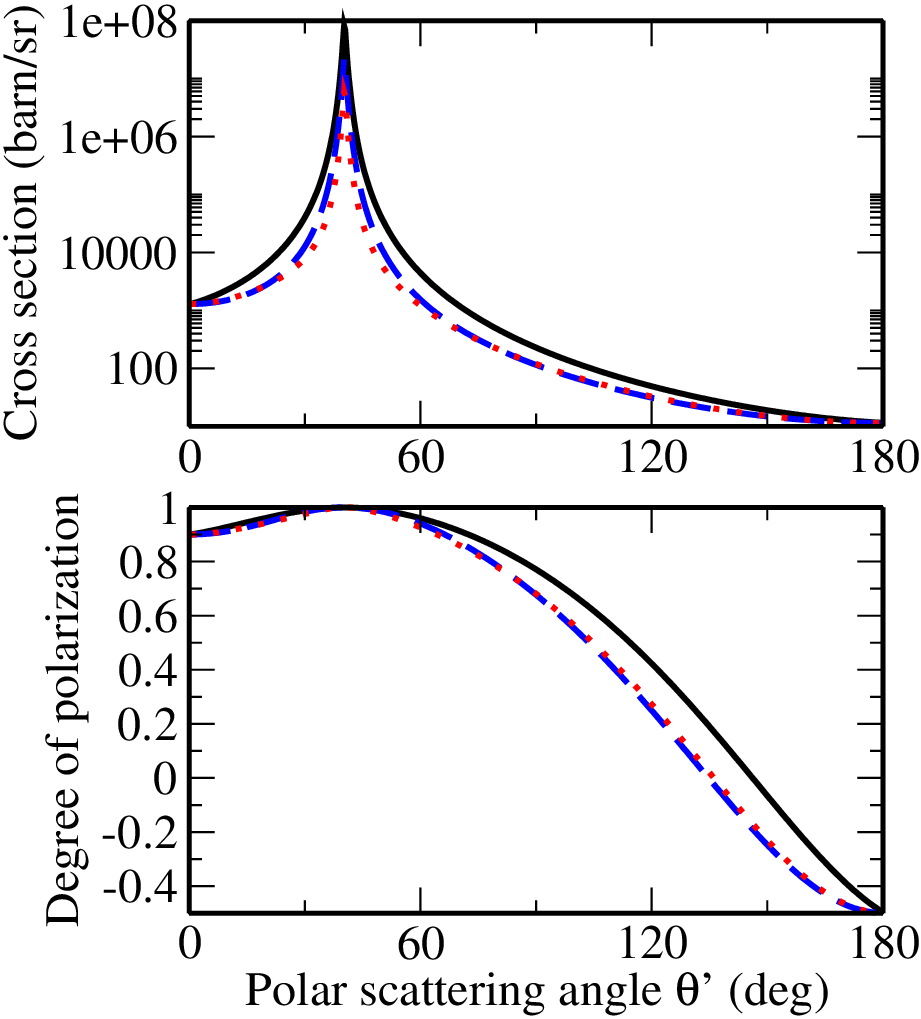}
  \caption{\label{Fig4}(Color online) The angle--differential cross section (upper panel) and the degree of longitudinal polarization (bottom panel) of the electrons scattered by an iron target. Calculations have been performed for a coherent superposition of two equally weighted Bessel beams with the difference of the TAM projections $\Delta m = 1$ (black solid line), $\Delta m = 2$ (blue dashed line) and $\Delta m = 3$ (red dotted line). The beams are prepared, moreover, in the state with the relative phase $\Delta \alpha$~=~60~deg, opening angle $\theta_p$~=~40~deg, helicity $\lambda = 1/2$, and kinetic energy $T_e = 300$~keV.}
\end{figure}

In the calculations above we have always assumed that the incident electrons are prepared in a pure quantum--mechanical state with the well--defined projection of the total angular momentum $m$ on the propagation ($z$--) axis. During the recent years, however, a number of experiments have been performed and/or proposed with the coherent superposition of two (or even more) states with different TAM projections \cite{MaH14,Iva12,GuS13}. For two Bessel beams with the same helicity $\lambda$ and the same beam axis, such a superposition is described by the wave--function:
\begin{equation}
   \label{eq_wave_function_twisted_electrons_superposition}
   \psi^{\rm (2 \, tw)}({\bm r}) = c_1 \psi_{\varkappa m_1 p_z \lambda}({\bm r}) + c_2 \psi_{\varkappa m_2 p_z \lambda}({\bm r}) \, ,
\end{equation}
where $\psi_{\varkappa m_n p_z \lambda}({\bm r})$ is given by Eq.~(\ref{eq_wave_function_twisted_electrons}) and $c_n$ coefficients read as:
\begin{equation}
   \label{eq_c_coefficients}
   c_n = |c_n| {\rm e}^{i \alpha_n} \, , \: \: \: |c_1|^2 + |c_2|^2 = 1 \, .
\end{equation}
With the help of this expression and Eq.~(\ref{eq_cross_section_twisted_averaged_1}) we can find the angle--differential cross section
\begin{eqnarray}
   \label{eq_cross_section_twisted_averaged_general_superposition}
   \frac{{\rm d}\sigma^{\rm (2 \, tw)}_{\lambda \lambda'}}{{\rm d}\Omega'}(\theta' , \, \theta_p) && \\
   && \hspace*{-2.0cm}=
   \frac{1}{\cos \theta_p} \, \int\limits_{0}^{2\pi} \left|f_{\lambda \lambda'}({\bm p}, {\bm p}') \right|^2 \, G(\varphi_p, \Delta m, \Delta \alpha) \frac{{\rm d}\varphi_p}{2 \pi} \,  \nonumber
\end{eqnarray}
for the Mott scattering of the (superposition of) Bessel beams by a macroscopic atomic target. In contrast to Eq.~(\ref{eq_cross_section_twisted_averaged_1}), derived for a single incident beam, this cross section depends on the function
\begin{equation}
   \label{eq_G_function}
    G(\varphi_p, \Delta m, \Delta \alpha) = 1 + 2|c_1||c_2| \, \cos \left(\Delta m (\varphi_p - \pi/2) +\Delta \alpha) \right) \, ,
\end{equation}
and, hence, on the differences of the TAM projections, $\Delta m = m_2 - m_1$, and phases, $\Delta \alpha = \alpha_2 - \alpha_1$, of two twisted states. Such a $\Delta m$-- (as well as $\Delta \alpha$-- ) dependence translates directly into the angular and polarization properties of scattered electrons. In Fig.~\ref{Fig4}, for example, we display the cross section ${\rm d}\sigma^{\rm (2 \, tw)}_{\lambda}/{\rm d}\Omega'(\theta', \, \theta_p) = \sum\limits_{\lambda'} {\rm d}\sigma^{\rm (2 \, tw)}_{\lambda \lambda'}/{\rm d}\Omega'(\theta', \, \theta_p)$ and the degree of longitudinal polarization $P^{\rm (2 \, tw)}(\theta; \theta_p)$ for the scattering of 300~keV electrons by the iron target. Calculations have been performed for the opening angle $\theta_p =$~40~deg, initial helicity $\lambda = 1/2$, and for $\Delta m = 1$ (solid line),  $\Delta m = 2$ (dashed line) and $\Delta m = 3$ (dotted line). As seen from the figure, both the cross sections and the polarization are strongly affected by the variation of the $\Delta m$. For example, the ${\rm d}\sigma^{\rm (2 \, tw)}_{\lambda}/{\rm d}\Omega'(\theta', \, \theta_p)$ decreases by about factor of 4 if observed at the angle $\theta' = \theta_p = 40$~deg and the difference of the TAM projections changes from $\Delta m = 1$ to $\Delta m = 3$. Such a remarkable effect can be easily observed by modern electron detectors and may provide useful information about the scattering of (superpositions of) Bessel electron beams.

%
%

%
%
%------------------------------------------------------  Summary and outlook  ---------------------------------------------------------------
%
%
\section{Summary and outlook}
\label{sec_summary}

The first Born approximation and Dirac's relativistic theory have been applied to explore the Mott scattering of high--energetic twisted electrons by atoms. In our study, we focused especially on the angular distribution and longitudinal polarization of outgoing electrons. To derive these---angular and polarization---properties we have approximated the atomic potential by (a sum of) Yukawa terms and considered two different ``experimental'' setups. In these setups, the incident Bessel beam collides with either (i) a single well--localized atom, or (ii) a macroscopic target, consisting of randomly distributed atoms. In the first case, we found that the angular distribution of the outgoing electrons depends not only on the kinematic parameters of the twisted wave, such as the ratio of the transverse $\varkappa$ to longitudinal $p_z$ momenta, but also on the (projection of the) total angular momentum $m$. The most pronounced effect of the TAM can be observed if the (single) target atom is placed close to the centre of the twisted wave front.

In contrast to the collision with a well--localized atom, the angle--differential cross section for the Mott scattering of a single Bessel wave by a \textit{macroscopic} target appears to be independent of the topological charge $m$. Nevertheless, the (angular and polarization) properties of the outgoing electrons are still sensitive for this case to the beam's opening angle, $\theta_p = \arctan (\varkappa/p_z)$. In particular, the angle $\theta_p$ determines the direction of the predominant emission of the outgoing electrons. Such an emission pattern, that is peaked at $\theta' = \theta_p$, provides a clear signature of the Mott scattering of twisted electrons and can be easily observed experimentally.

For a macroscopic target we have also analyzed the scenario in which the incident electron beam is prepared as a coherent superposition of \textit{two} Bessel states. The use of such a superimposed beam helps to restore the sensitivity of the Mott process on the projections $m$ of the TAM. In particular, our calculations clearly indicate that the angular and polarization properties of the scattered electrons strongly depend on the difference $m_1 - m_2$. This $\Delta m$--dependence can be observed most clearly at the scattering angle $\theta' = \theta_p$ at which both the angle--differential Mott cross section and the longitudinal polarization of scattered electrons are maximal.

The present theoretical study has been performed in the (relativistic) first Born approximation. While the accuracy of this approach is limited, especially for relatively low collision energies and backward scattering angles, it helps to elucidate the main features of the elastic scattering of twisted electrons. Moreover, based on the developed theory, one can perform a more elaborate analysis of the Mott process, including the spin--interaction effects and higher perturbation terms. Such an analysis is currently underway  and its results will be published elsewhere.

%
%
% -----------------------------------------Acknowledgements---------------------------------------------------------
%
%
\section*{Acknowledgements}

We are grateful to G. Kotkin and D. Karlovets for useful discussions. V.~G.~S. acknowledges support from RFBR (via grant No. 13--02--00695) and MES (Russia).

%
%
% ---------------------------------------------------- Appendix -----------------------------------------------
%
%
\begin{appendix}
\section{Twisted electron wave--function}
\label{appendix_wave_function}

Apart from this study, the wave--function of relativistic twisted electrons has been utilized in several recent works \cite{BlD11,Kar12,SeS14}. Since the explicit form of this function slightly differs in each paper, let us discuss and compare here the various representations of corresponding wave-functions. We start from our definition (\ref{eq_wave_function_twisted_electrons}) which, upon substitution of the plane--wave solution (\ref{eq_plane_wave_initial}) and the amplitude (\ref{eq_a_amplitude}), reads as:
\begin{eqnarray}
   \label{eq_wave_function_twisted_appendix}
   \psi_{\varkappa m p_z \lambda}({\bm r}) &=& \sum\limits_{\sigma} \int{\frac{{\rm d} \varphi_p}{2 \pi}} \, \sqrt{\frac{\varkappa}{2 \pi}}
   \, {\rm e}^{i (p_z z + \varkappa r_{\perp} \cos(\varphi_p - \varphi_r))} \nonumber \\[0.3cm] \nonumber \\
   && \hspace*{-1.3cm} \times (-i)^m \, {\rm e}^{i (m - \sigma) \varphi_p} \, d^{1/2}_{\sigma \lambda}(\theta_p)\, U^{(\sigma)}(\varepsilon, \lambda) \, ,
\end{eqnarray}
where we performed trivial integration over $p_\perp$ and, for the sake of shortness, introduced the notation:
\begin{eqnarray}
   \label{eq_U_spinor}
   U^{(\sigma)}(\varepsilon, \lambda) = \left( \begin{array}{c}
                                   \sqrt{\varepsilon + m_e} \, w^{(\sigma)}({\bm e}_z) \\[0.4cm]
                                   2 \lambda \sqrt{\varepsilon - m_e} \, w^{(\sigma)}({\bm e}_z)
                                   \end{array}
                          \right) \, .
\end{eqnarray}
This bi--spinor $U^{(\sigma)}(\varepsilon, \lambda)$ is obviously the eigensolution of the spin operator
\begin{equation}
   \Sigma_z = \frac{1}{2} \, \left(\twobytwo{\sigma_z}{0}{0}{\sigma_z} \right) \, ,
\end{equation}
with the eigenvalues $\sigma = \pm 1/2$.

By using the well--known integral relation
\begin{equation}
   \frac{1}{2\pi} \, \int_0^{2\pi} \, {\rm e}^{ i n \varphi + i z \cos{\varphi}} \, {\rm d}\varphi = i^n \, J_n(z) \, ,
\end{equation}
we can perform the integration over the azimuthal angle $\varphi_p$ in Eq.~(\ref{eq_wave_function_twisted_appendix}) and obtain the twisted wave--function:
\begin{eqnarray}
   \label{eq_wave_function_twisted_appendix_2}
   \psi_{\varkappa m p_z \lambda}({\bm r}) &=& \sqrt{\frac{\varkappa}{2 \pi}} \, {\rm e}^{i p_z z} \,
   \sum\limits_{\sigma} i^{-\sigma} \, {\rm e}^{i (m - \sigma) \varphi_r} \, d^{1/2}_{\sigma \lambda}(\theta_p) \nonumber \\
   &\times& J_{m - \sigma}(\varkappa r_\perp) \, U^{(\sigma)}(\varepsilon, \lambda) \, ,
\end{eqnarray}
in cylindrical coordinates $(r_\perp, \varphi_r, z)$. 
The normalization condition for these states is

 \begin{eqnarray}
\int \bar \psi_{\varkappa m p_z \lambda}({\bf r})\,
\psi_{\varkappa' m' p_z' \lambda'}({\bf r})\, d^3 r&=&
\delta(\varkappa-\varkappa')\,\delta_{mm'}
 \\
&\times& 2\pi\delta(p_z-p_z') \cdot 2m_e\delta_{\lambda\lambda'}\,.
\nonumber
 \end{eqnarray}
Each summand in the right--hand side of Eq.~(\ref{eq_wave_function_twisted_appendix_2}) is the eigen--function of the operators $\hat{L}_z = -i \partial/\partial\varphi_r$, $\Sigma_z$ and $\hat{J}_z = \hat{L}_z + \hat{\Sigma}_z$ with eigen--values $m - \sigma$, $\sigma$ and $m$, respectively. However, due to the summation over $\sigma = \pm 1/2$ in Eq.~(\ref{eq_wave_function_twisted_appendix_2}), the $\psi_{\varkappa m p_z \lambda}({\bm r})$ does not possess definite ($z$--~) projections of the spin and orbital angular momentum separately. It is the eigen--state of the total angular momentum operator, $\hat{J}_z \psi_{\varkappa m p_z \lambda}({\bm r}) = m \psi_{\varkappa m p_z \lambda}({\bm r})$, with \textit{half--integer} $m$. Therefore, the Bessel vortex state (\ref{eq_wave_function_twisted_appendix_2}) (or (\ref{eq_wave_function_twisted_appendix})) of free relativistic electrons includes intrinsically spin--orbit interaction \cite{BlD11}.

The coordinate representation of the wave--function (\ref{eq_wave_function_twisted_appendix_2}) can be easily generalized to the case when the vortex electron beam is shifted by the vector ${\bm b} = (b_x, b_y, 0)$ with respect to the quantization $z$--axis. Inserting the translation factor ${\rm exp}(-i {\bm p} {\bm b})$ into Eq.~(\ref{eq_wave_function_twisted_electrons}) and performing integration over the ${\bm p}_\perp$, we find:
\begin{eqnarray}
   \label{eq_wave_function_twisted_appendix_3}
   \psi_{\varkappa m p_z \lambda}({\bm r}, {\bm b}) &=& \sqrt{\frac{\varkappa}{2 \pi}} \, {\rm e}^{i p_z z} \,
   \sum\limits_{\sigma} i^{-\sigma} \, {\rm e}^{i (m - \sigma) \varphi_\rho} \, d^{1/2}_{\sigma \lambda}(\theta_p) \nonumber \\
   &\times& J_{m - \sigma}(\varkappa \rho_\perp) \, U^{(\sigma)}(\varepsilon, \lambda) \, ,
\end{eqnarray}
where ${\bm \rho} = {\bm r}_{\perp} - {\bm b} = (\rho_\perp, \varphi_\rho, 0)$ is the displacement vector in the $x-y$ plane written in cylindrical coordinates. An alternative representation of the (displaced) Bessel electron wave--function was proposed in the non--relativistic framework in Refs.~\cite{VaB13,YuL13}. It relies on the addition theorem for Bessel functions and, although being mathematically equivalent to Eq.~(\ref{eq_wave_function_twisted_appendix_3}), is less convenient for practical purposes and does not allow a transparent physical interpretation.

By making use of Eqs.~(\ref{eq_wave_function_twisted_appendix_2}) and (\ref{eq_wave_function_twisted_appendix_3}), it is easy to investigate the behaviour of the Bessel electron wave--function for vanishing values of the transverse momentum, $\varkappa \to 0$, and, hence, the opening angle $\theta_p \to 0$. In this limit $d^{1/2}_{\sigma \lambda}(\theta_p) = \delta_{\sigma \lambda}$ and $J_{m - \sigma}(\varkappa \rho_\perp) = \delta_{m - \sigma, 0}$, and, hence,
\begin{eqnarray}
   \label{eq_wave_function_twisted_appendix_4}
   \frac{\psi_{\varkappa m p_z \lambda}({\bm r}, {\bm b})}{\sqrt{\varkappa}} \Bigg|_{\varkappa \to 0} &=&
   \frac{i^{-\lambda}}{\sqrt{2 \pi}} \, \, u_{{\bm p} \lambda} \, {\rm e}^{i {\bm p} {\bm r}} \, \, \delta_{m = \lambda} \, ,
\end{eqnarray}
where the $u_{{\bm p} \lambda}$ is the standard Dirac bi--spinor (\ref{eq_Dirac_bispinor}), and ${\bm p} = (0, 0, p_z)$. As can be seen from this expression, the Bessel electron wave function recovers---for $\varkappa \to 0$---the standard solution for a plane wave that propagates along the $z$--axis with the helicity $\lambda = m = \pm 1/2$.

As mentioned already above, some another representation of the wave--function $\psi_{\varkappa m p_z \lambda}({\bm r})$ has been proposed recently
by Karlovets \cite{Kar12} which reads as:
\begin{eqnarray}
   \label{eq_wave_function_twisted_appendix_Karlovets}
    \widetilde{\psi}_{\varkappa m' p_z \lambda}({\bm r}) &=& {\rm e}^{i \lambda \varphi_r} \sqrt{\frac{\varkappa}{2 \pi}} \, {\rm e}^{i p_z z} \,
   \sum\limits_{\sigma} {\rm e}^{i (m' - \sigma) \varphi_r} \, d^{1/2}_{\sigma \lambda}(\theta_p) \nonumber \\
   &\times& J_{m' - \sigma}(\varkappa r_\perp) \, U^{(\sigma)}(\varepsilon, \lambda) \, ,
\end{eqnarray}
which differs from Eq.~(\ref{eq_wave_function_twisted_appendix_2}) by (i) the overall pre--factor ${\rm exp}(i \lambda \varphi_r)$ and (ii) the missing term $i^{-\sigma}$ under the summation sign. The first factor leads to the fact that the $\widetilde{\psi}_{\varkappa m' p_z \lambda}({\bm r})$ is the eigen--function of the total angular momentum operator $\hat{J}_z$ with the eigen--value $m' + \lambda$, where $m'$ is the \textit{integer}. This factor comes just from the different convention in the definition of the free--electron spinor, $\widetilde{w}^{(\lambda)}({\bm n}) = {\rm exp}(i \lambda \varphi) w^{(\lambda)}({\bm n})$, where $w^{(\lambda)}({\bm n})$ is given by (\ref{eq_Dirac_spinor_arbitrary_direction}), and does not affect---upon relabeling of the quantum numbers---the Mott scattering cross section. In contrast, the term $i^{-\sigma}$ which defines the relative phase of two components of the wave--function (\ref{eq_wave_function_twisted_appendix_Karlovets}) with spin projections $\sigma = \pm 1/2$, was mistakenly missed in Ref.~\cite{Kar12}. Without including this factor, the $\widetilde{\psi}_{\varkappa m' p_z \lambda}({\bm r})$ is \textit{not} a solution of the Dirac equation with a definite energy.

Yet another representation of the Bessel electron wave--function was given by Bliokh and co--workers in Ref.~\cite{BlD11}. For the sake of shortness, we will not present here this---rather lengthy---expression. We just mention that instead of our Eq.~(\ref{eq_wave_function_twisted_appendix_3}), constructed from the (plane--wave) components (\ref{eq_plane_wave_functions})--(\ref{eq_Dirac_bispinor}) with a definite \textit{helicity} $\lambda$, Ref.~\cite{BlD11} employs the plane--waves whose polarization state is defined with respect to some (overall) quantization axis in the electron rest frame:
\begin{equation}
   \label{eq_plane_wave_Bliokh}
   \tilde{\psi}_{{\bm p}}({\bm r}) = {\rm e}^{i {\bm p} {\bm r}} \,
                         \left( \begin{array}{c}
                                   \sqrt{\varepsilon + m_e} \, w^{(\lambda)}({\bm e}_z) \\[0.4cm]
                                   \sqrt{\varepsilon - m_e} \, \left( {\bm \sigma} {\bm n} \right)\, w^{(\lambda)}({\bm e}_z)
                                   \end{array}
                          \right) \, .
\end{equation}
Although the application of Eq.~(\ref{eq_plane_wave_Bliokh}) is useful to explore the spin and orbital angular momentum currents in free--propagating vortex beams, it is less convenient for the analysis of high--energy electron scattering.

\end{appendix}

%
%
%--------------------------------------Bibliography-----------------------------------------------------------------
%
%

\end{document}